\documentclass[aps,pre,groupedaddress,onecolumn]{revtex4}

\usepackage{amsmath}
\usepackage{amssymb}
\usepackage{graphicx}
\usepackage{dcolumn}
\usepackage{hyperref}
\usepackage{bm}
\usepackage[usenames,dvipsnames]{xcolor}

\newcommand\We{\textrm{We}}
\newcommand\Rey{\textrm{Re}}

\newcommand\di{\textrm{d}}
\newcommand\mN{\mathcal{N}}

\newcommand\avg[1]{#1}

\newcommand{\al}[1]{{#1}}

\begin{document}
\title{Capillary driven fragmentation of large gas bubbles in turbulence}
\author{Ali\'enor Rivi\`ere$^{1,2,3}$, Daniel J. Ruth$^{1}$, Wouter Mostert$^{1,4}$, Luc Deike$^{1,5}$ and St\'ephane Perrard$^{2,3}$ \thanks{Email address for correspondence: stephane.perrard@espci.fr}}
\affiliation{$^1$Department of Mechanical and Aerospace Engineering, Princeton University, Princeton, NJ 08544, United States\\
$^2$LPENS, D\'epartement de Physique, Ecole Normale Sup\'erieure, PSL University, 75005 Paris, France, EU.\\
$^3$ Physique et M\'ecanique des Milieux H\'et\'erog\`enes, CNRS, ESPCI Paris, University PSL, Paris 75005, France, EU.
$^4$Department of Mechanical and Aerospace Engineering, Missouri University of Science and Technology, Rolla, MO 65401, United States \\
$^5$High Meadows Environmental Institute, Princeton University, Princeton, NJ 08544, USA
}

\begin{abstract}
The bubble size distribution below a breaking wave is of paramount interest when quantifying mass exchanges between the atmosphere and oceans. Mass fluxes at the interface are driven by bubbles that are small compared to the Hinze scale $d_h$, the critical size below which bubbles are stable, even though individually these are negligible in volume. Combining experimental and numerical approaches, we report a power law scaling $d^{-3/2}$ for the small bubble size distribution, for sufficiently large separation of scales between the injection size and the Hinze scale. From an analysis of individual bubble break-up events, we show that break-ups generating small bubbles are driven by capillary effects, and that their break-up time scales as $d^{3/2}$, which physically explains the sub-Hinze scaling observed.
\end{abstract}

\pacs{??}

\maketitle

Bubble fragmentation drives gas dissolution by drastically increasing the exchange surface between phases. For instance, up to 40\% of the total CO$_2$ uptake by the ocean is due to bubble-mediated gas transfer~\citep{Keeling1993,Deike2018gas,Reichl2020}. More specifically, it is the bubble size distribution that controls gas transfer \citep{leighton2018,atamanchuk2020} and spray production as bubbles burst at the surface~\cite{spiel1996,ghabache2014,Deike_PRF_2018,berny2020}. Bubbles also play a major role in industrial applications like oil and gas transportation from remote wells~\citep{Galinat2005} or oil spill mitigations~\citep{Katz2010,oilspill1}.

As a consequence, the fragmentation of bubbles has been extensively studied in model experiments~\cite{Risso1998,Lasheras1999,lalanne2019model,Qi2020} as well as under breaking waves both experimentally~\citep{Loewen_1996,Deane_2002,Rojas_2007,blenkinsopp2010} and numerically~\cite{Deike_JFM_2016,Mostert_2021,Chan_2021}. For large bubbles, a consensus has been reached on the bubble size distribution, described as $\mN(d) \propto d^{-10/3}$ with $d$ the bubble volume equivalent diameter. This law originates from a self-similar cascade of break-ups~\cite{Garrett2000}, in which each bubble produces a fixed number of equally sized child bubbles, on a time given by the turbulent flow correlation time at the bubble scale $d$. This time is the eddy turnover time $t_c(d) = \epsilon^{-1/3} d^{2/3}$ where $\epsilon$ is the averaged dissipation rate of kinetic energy by viscous dissipation, and which is commonly used to describe turbulent flows \cite{pope2000}. 

The $\mN(d) \propto d^{-10/3}$ scaling holds \al{down to the Hinze scale $d_h$,} the size at which kinetic energy balances surface tension energy. The ratio of inertial and surface tension effects defines the Weber number, $\We=\rho U^2 d/\gamma$, where $\gamma$ is the liquid-gas surface tension, $\rho$ the liquid density, and $U$ a characteristic velocity driving the interface deformation. For a bubble immersed within a turbulent flow, the characteristic velocity $U$ is the averaged velocity difference across the scale of the bubble. If the bubble is within the inertial range, {\it i.e.} where the turblence is scale invariant, the velocity increment is given by $U^2= 2 \epsilon^{2/3}d^{2/3}$ \cite{K41}. This yields the Weber number for a bubble in a turbulent flow,
\begin{equation}
\We = \frac{2 \rho \epsilon^{2/3} d^{5/3}}{\gamma}. 
\end{equation}
Concomitantly, the Hinze scale $d_h$~\cite{Hinze_1955} sets the limit between bubbles that will fragment ($d>d_h$) and stable bubbles ($d<d_h$) with $\We_c$ the critical Weber number for break-up, and reads
\begin{equation}
d_h = \left (\frac{\We_c}{2}\right )^{3/5} \left (\frac{\gamma}{\rho} \right )^{3/5} \epsilon^{-2/5}.
\end{equation}
Note that due to the inherent stochasticity of turbulent flows, $d_h$ is a soft break-up limit and not necessarily a fixed constant, and will depend on the observation time.

For its part, the sub-Hinze bubble size distribution ($d\ll d_h$) always exhibits a gentler slope than $\mN(d) \propto d^{-10/3}$, although there is variability among the experimental studies~\cite{Loewen_1996,Deane_2002,Rojas_2007,blenkinsopp2010}, and the observations lack a physical explanation~\cite{Deane_2002}. The difficulty arises from the large scale separation between sub-Hinze bubbles and their parent size bubbles: the sub-Hinze bubble distribution cannot be explained by a self-similar cascade process, so a different type of physical argument is required.


\begin{figure*}
\centering
\includegraphics[width=0.78\textwidth]{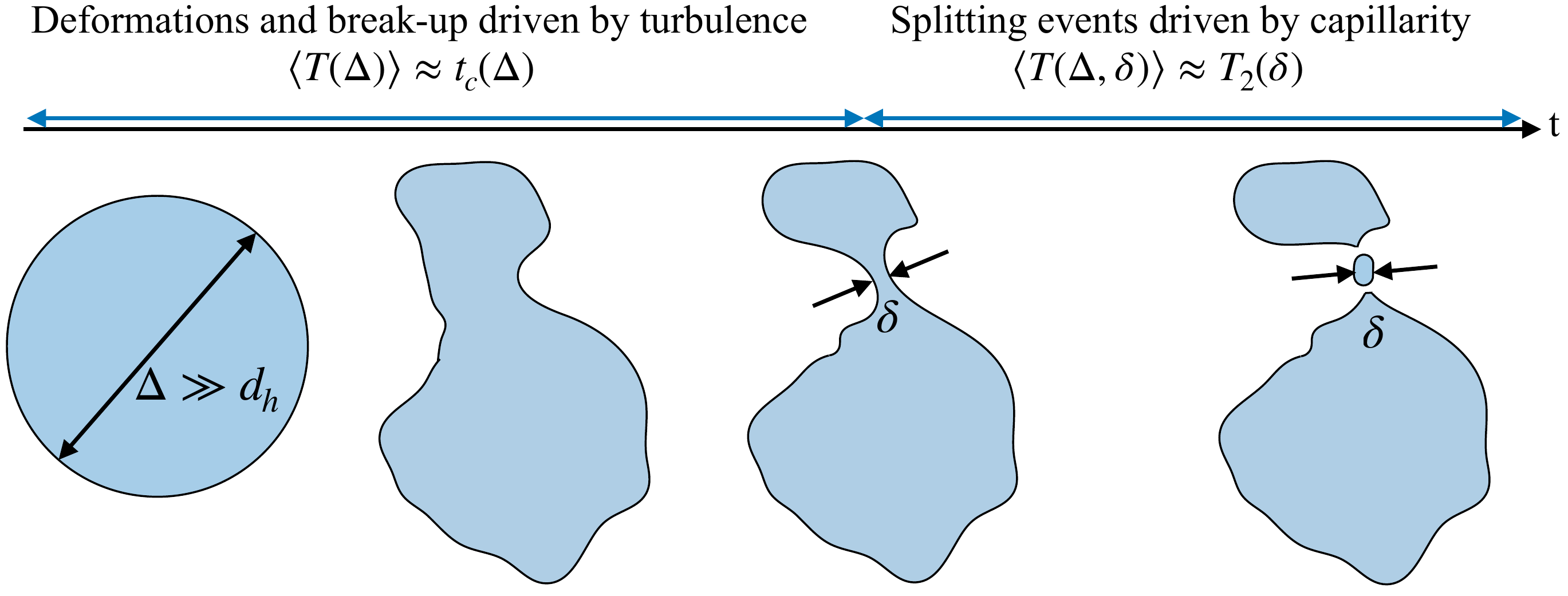}
\includegraphics[width=0.2\textwidth]{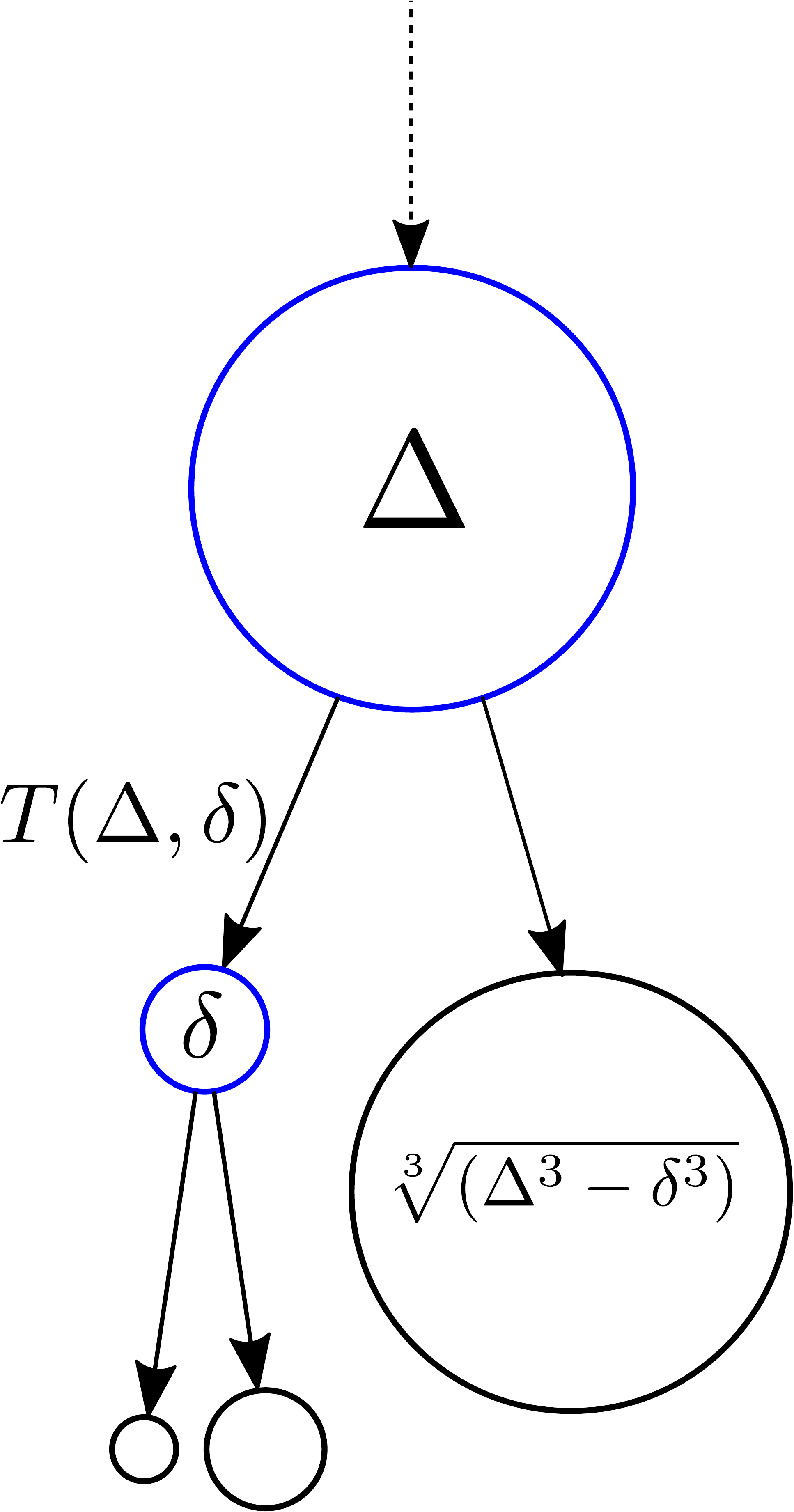}
\caption{(a) Sketch of a break-up process, involving inertial deformations followed by capillary splitting events. (b) Schematic of a bubble of size $\Delta$ splitting into a small bubble of size $\delta$ in a time $T(\Delta, \delta)$ and subsequent splittings.}
\label{sketch}
\end{figure*}

In this letter, we introduce a new physical ingredient to explain the origin of the sub-Hinze bubble distribution. \al{We decompose breaking sequences into binary events and show that the bubble fragmentation dynamics of large super-Hinze bubbles result from two concomitant processes: {\it break-up events}, which lead to the formation of two super-Hinze bubbles, and {\it splitting events}, which create at least one sub-Hinze bubble. Break-up events are self-similar events occurring on the eddy turn-over time at the parent bubble scale. Splitting events happen much faster and are capillary driven. They take place on filaments structures that have been prepared by break-up events. Figure \ref{sketch} sketches these processes.} Using a population balance equation that considers these two types of processes, we reproduce the two power law scalings for the bubble size distribution for $d>d_h$ and $d<d_h$ respectively. We then use three-dimensional two-phase direct numerical simulations (DNS) of bubble break-up in turbulence to validate the population model ingredients, by analyzing the statistics of individual splitting events. \al{Finally, we present an experiment that achieves a large scale separation between an initial large bubble and the Hinze scale, and which produces} a clear sub-Hinze bubble size distribution power law $\mathcal{N}(d) \propto d^{-3/2}$, in accordance with the theoretical model and the numerical simulations.


We consider the fragmentation dynamics as a succession of binary events. Let $T(\Delta,\delta)$ denote the lifetime of the parent bubble, of equivalent diameter $\Delta$, which produces two child bubbles of equivalent diameters $\delta$ and $\sqrt[3]{\Delta^3-\delta^3} \geq \delta$ (from volume conservation), as sketched in figure~\ref{sketch}. For equal-size child bubbles we have $\delta = c \Delta$ with $c = 2^{-1/3}\approx 0.79$. \al{Classically, the production rate of a bubble of size $d$ from the break-up of a bubble of size $\Delta$ can be decomposed into the product of a break-up rate $\omega(\Delta)$ and a child size probability density $f(\Delta, d)$~\cite[chapter~4, equation~2.1]{ethier2009}. $f$ is often referred to as the child bubble size distribution and is the probability density function for a child of size $d$ given the break-up of a bubble of size $\Delta$. The mean bubble flux $\Phi(\Delta, d, t)$ from size $\Delta$ to $d$ can be written as
\begin{equation}
\Phi(\Delta,d, t) = 2f(\Delta, d) \omega(\Delta) \mathcal{N}(\Delta,t),
\label{flux}
\end{equation}
where $\mathcal{N}(\Delta, t)$ is the number density of bubbles of size $\Delta$ at time $t$ and the numerical factor comes from the fact that every event is binary \cite{Lasheras1999,Qi2020}. From Eq.~\ref{flux}, we obtain the temporal evolution of the bubble size distribution from the fragmentation of an initial bubble of size $d_0$ \cite{Lasheras1999}:
\begin{equation}
\frac{\di \mathcal{N}(d,t)}{\di t} = \int_{d}^{d_0} \Phi(\Delta,d,t) \di \Delta - \omega(d)\mN(d),
\label{fullbudget}
\end{equation}}
written down as a difference between a birth term and a death term. Equation \eqref{fullbudget} is the starting point of general population balance models. For super-Hinze bubbles $d>d_h$ and considering break-up rates controlled by the eddy turn-over time at the scale of the parent bubble, using \eqref{fullbudget} at steady state one recovers the $\mathcal{N}(d)\propto d^{-10/3}$ regime \citep{Garrett2000,Chan_2021}. 
\al{Indeed, in this self similar model a bubble of size $d$ breaks, in a time given by $t_c(d)=\epsilon^{-1/3} d^{2/3}$, into $m$ fragments of equal diameter $m^{-1/3} d$. Each of these child bubbles then breaks in a time $\epsilon^{-1/3} (m^{-1/3} d)^{2/3}$. We obtain an increasing number of $m \times m ^{-2/9} = m^{7/9}$ bubbles per unit of time, which yields the bubble density $\mathcal{N}(m^{-1/3}d) = m^{1/3} m^{7/9} \mathcal{N}(d)$. Assuming $\mathcal{N}(d) \propto d^{\alpha}$, we have $-\alpha/3   = 1/3+7/9$ which gives $\alpha = -10/3$. For $d < d_h$, the self-similar argument cannot be applied anymore since surface tension must be important at this scale.}

Here we aim to obtain a scaling for the sub-Hinze bubble size distribution. In equation \eqref{flux} the rate $\omega(\Delta)$ at which a bubble breaks up does not distinguish between processes which produce equally sized child bubbles or highly asymmetrically sized child bubbles, \al{for which at least one child bubble is smaller than the Hinze scale}. These two types of events, however, may occur on very different timescales. Here we assume that the production of small bubbles ($d < d_h$) is controlled by bubble splitting events, in which elongated filaments become unstable under a Rayleigh-Plateau-like mechanism~\cite{Villermaux_2020}. As such, the time for such elongated filaments to rupture will be controlled by capillarity, at the scale of the filament. \al{This stems from the "freezing" in place of the turbulent flow relative to the accelerating collapse dynamics in the final moments before rupture, which was shown experimentally in \citet{Ruth_PNAS_2019}}. A bubble splitting event leads to the formation of \al{at least one sub-Hinze bubble, whose size $\delta$ is comparable to the diameter of the filament, and one larger bubble of size $\sqrt[3]{\Delta^3-\delta^3}$}. The exact geometry of the filament and the splitting time varies from one realization to the other, but under an ensemble average, the splitting time $\avg{T(\Delta,\delta)}$ will be given by the capillary time at size small $\delta$:
\begin{equation}
\langle T(\Delta,\delta) \rangle =   T_2(\delta) = \frac{1}{2\sqrt{3}}  \left ( \frac{\rho}{\gamma} \right )^{1/2} \delta^{3/2}.
\label{eqtime}
\end{equation}
Note that this time depends only on the size of the child bubble, not on the parent. To integrate the splitting events within the population balance framework, we use flux conservation to express $\omega$ and $f$ from equation \eqref{flux} in terms of the newly introduced timescale:
\al{
\begin{equation}
f(\Delta,d) \omega(\Delta) =  \frac{F(\Delta, \delta)}{\langle T(\Delta, \delta)\rangle} = \frac{F(\Delta, \delta)}{T_2(\delta)},
\label{eqkernel}
\end{equation}
with $\delta=d$ if the child bubble considered is the smaller one of the two produced (that is, if $d<c\Delta$ ) and $\delta = \sqrt[3]{\Delta^3 - d^3}$ if it is the larger of the two (that is, if $d>c\Delta$), since the production is controlled by the faster of the two timescales}. $F(\Delta, \delta)$ is the weight associated to each break-up frequency. We will use DNS of bubble break-up in homogeneous and isotropic turbulence to estimate $F$ and will find that $F(\Delta, \delta)\equiv F(\Delta)$, independent of $\delta$. We will work with this assumption in the remaining of the theoretical discussion.

\al{In this framework, using equation~\eqref{eqkernel}, equation~\eqref{fullbudget} reads:
\begin{equation}
\frac{\di \mathcal{N}(d,t)}{\di t} = \int_{d/c}^{d_0}  ~\al{2} \frac{F(\Delta)}{T_2(d)} \mathcal{N}(\Delta,t)~\di \Delta + \int_{d}^{d/c}  ~\al{2} \frac{F(\Delta)}{T_2(\sqrt[3]{\Delta^3 - d^3})} \mathcal{N}(\Delta,t)~\di \Delta - \omega(d)\mN(d).
\end{equation}
We assume that sub-Hinze bubbles do not break. For bubbles such that $d<cd_h$, this implies that the second integral and the death term vanish and that the lower bound of the first integral is $d_h$ leading to: }
\begin{equation}
\frac{\di \mathcal{N}(d,t)}{\di t} = \int_{d_h}^{d_0}  ~\al{2} \frac{F(\Delta)}{T_2(d)} \mathcal{N}(\Delta,t)~\di \Delta
\label{budget}
\end{equation}
Integrating over time, we obtain for \al{$d<cd_h$}
\begin{eqnarray}
\mathcal{N}(d,t) =  d^{-3/2} \int_0^t I_\mathcal{N}(d_0/d_h,s) \di s,
\end{eqnarray}
with,
\al{
\begin{eqnarray}
I_\mathcal{N}(d_0/d_h,t) = \int_{d_h}^{d_0}  ~ 4\sqrt{3} F(\Delta) \left ( \frac{\rho}{\gamma} \right )^{-1/2} \mathcal{N}(\Delta,t)\di \Delta.
\end{eqnarray}
}
The integral $I_\mathcal{N}$ does not depend on the child bubble size $d$, so that the bubble size distribution for \al{$d<cd_h$} follows 
\begin{eqnarray}
\mathcal{N}(d,t) \propto d^{-3/2}, 
\end{eqnarray}
with the scaling exponent $-3/2$ independent of the details of the break-up cascade above the Hinze scale.\\

		\begin{figure}
\centering
\includegraphics[width=0.8\columnwidth]{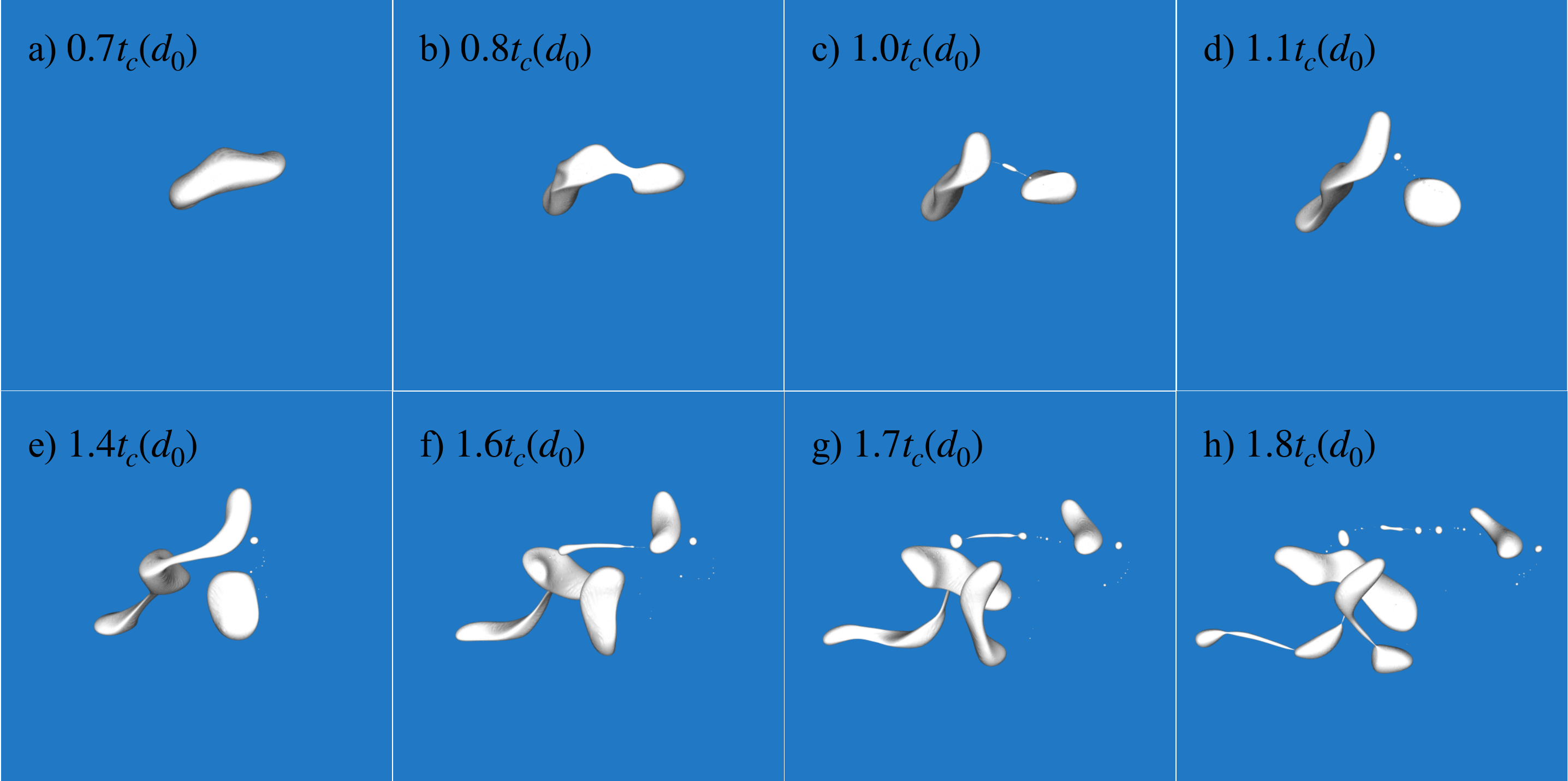}
\caption{DNS snapshots of a typical break-up sequence, with the initial bubble size $d_0/d_h=2.9$. The bubbles' interface is represented in white. The first images (a,b) show large scale deformation due to turbulence, happening over the eddy turn over time at the size of the initial bubble scale $d_0$, $t_c(d_0) = \epsilon^{-1/3}d_0^{2/3}$, leading to the formation of thin filaments (c,d). Successive splitting events of the filament are visible (e,f,g,h) leading to multiple child bubbles. The filaments quickly break creating a wide range of bubble sizes, the smallest being orders of magnitude smaller than the initial one. }
\label{fig:simus}
\end{figure} 

	To evaluate the validity of the physical arguments leading to $\mN(d, t) \propto d^{-3/2}$, we perform DNS of a single initial bubble larger than the Hinze scale in a turbulent flow using the free software Basilisk~\citep{Popinet_2009,Popinet_2018}. A detailed description of the numerical configuration can be found in \citet{Riviere2021}. We first create a homogeneous and isotropic turbulent flow at Taylor Reynolds number $\Rey_\lambda=38$, following the method introduced by~\citet{Meneveau_2005}.  We then introduce a spherical bubble of diameter $d_0$ within the inertial range of the Kolmogorov cascade~\citep{K41}, i.e. at a scale where turbulence is scale invariant. We vary the ratio $d_0/d_h$ and we have verified that the velocity statistics at the scale of the parent bubble are typical of turbulent flows~\cite{Perrard_JFM_2020} (although the Taylor Reynolds number is smaller than that in typical experimental conditions). We perform at least ten simulations per value of the initial bubble size $d_0/d_h$ (2.9, 4.1, 5.2) with a spatial resolution of 135 points per diameter.
	
	Figure~\ref{fig:simus} presents snapshots of a large bubble (giving an initial separation of scales $d_0/d_h=2.9$) subject to large deformations, described in detail in \citet{Riviere2021}. The initial break-up, which occurs within one eddy turnover time at the bubble scale $t_c(d_0)$~\citep{Riviere2021}, is followed by a rapid succession of splitting events, occurring on a much faster time scale and producing dozens of sub-Hinze scale bubbles. 

	\al{As previously, }we decompose the dynamics into binary events and associate a lifetime $T(\Delta,\delta)$ to each parent bubble of size $\Delta$ producing a small child bubble of size $\delta$. We compute the values of the equivalent diameters $\Delta$ and $\delta$ from parent and child bubble volumes. All individual bubbles are tracked from birth to death to determine $T(\Delta,\delta)$ using a reconstruction process of the full event sequence for each simulation. To do so, all individual bubbles are first tracked in space and time using the Python package trackpy~\citep{trackpy_v4} based on the Crocker-Grier algorithm~\cite{crocker1996}. Using volume and momentum conservation during break-up events, we reconstruct the breakage tree event by event. Each criterion has been manually adjusted and tested on simple situations to validate the algorithm robustness. The processing is systematically applied to the entire data set, and leads to the identification of 4329 breaking events for $d_0/d_h$ ranging from 2.9 to 5.2, using 78 different 3D DNS realizations of bubble break-up. In the following, we focus on the sub-Hinze bubble production, corresponding to $\delta<d_h$ and $t< 4 t_c(d_0)$, during which most of the sub-Hinze bubbles are generated~\cite{Riviere2021}. Given the low volume fraction of air, the coalescence events are statistically negligible. 
	
\begin{figure}
\centering
\includegraphics[width=0.6\textwidth]{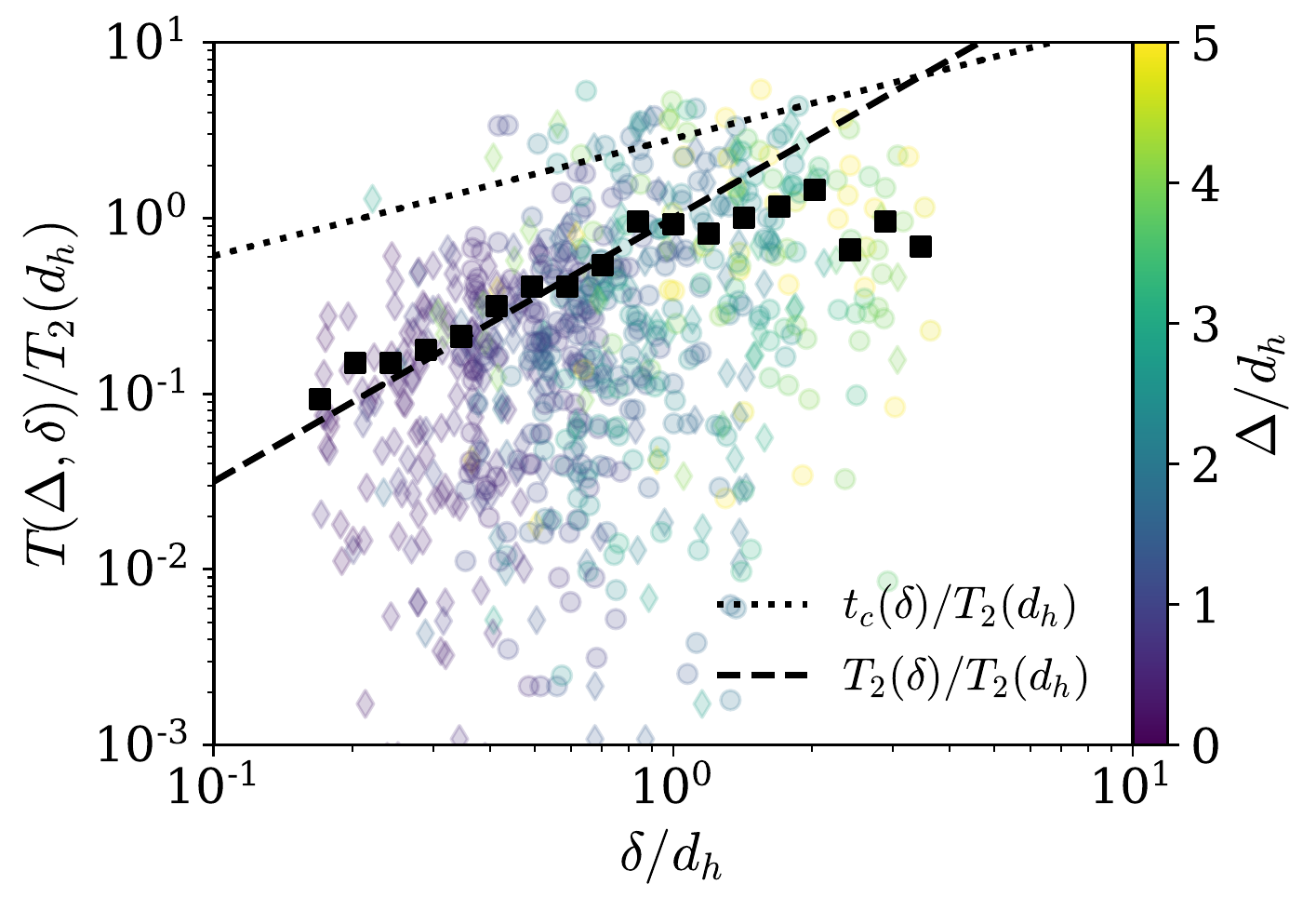}
\caption{Lifetime of bubbles as a function of the size of the smallest bubble they split apart into. The size of the parent bubble is given by the color. All splitting events occur on a time scale faster than the eddy turn-over time $t_c(\delta)$, shown in dotted line, suggesting that the splitting events are not instigated by turbulent deformations at the small child bubble size. The ensemble averaged time $\langle T(\Delta, \delta)\rangle_\Delta$ is shown in black squares and follows the capillary timescale of the small child bubble $T_2(\delta)$ without any adjusting parameters (shown in dashed line).} 
\label{fig:pdf-T}
\end{figure}

	Figure \ref{fig:pdf-T}a shows the splitting times $ T(\Delta,\delta)$ as a function of the size of the smallest child bubble they produce, $\delta$. Each individual event is color coded by the parent size $\Delta$, highlighting a broad distribution of splitting times, all smaller than the eddy turn-over time at this small child bubble's scale, $t_c(\delta)$. This suggests that these splitting events are not instigated by turbulent deformations at the small child scale. 
To estimate the ensemble average $\langle T(\Delta, \delta)\rangle$ over multiple realizations, we compute the ensemble average over $\Delta$-values given $\delta$, denoted $\langle T(\Delta, \delta)\rangle_\Delta$, which is shown in black squares. It clearly matches the capillary time scale $T_2(\delta)$ \textit{i.e.} the typical capillary time at the length $\delta$ (shown in black dashed line), up to $\delta = d_h$, without any adjustable parameters. For $\delta>d_h$, the break-up time seems to converge to a value independent of $\delta$. We have also separately verified that, $\langle T(\Delta, \delta) \rangle$ being a function of two variables, the ensemble average of over $\delta$ values for a given $\Delta$, $\langle T(\Delta, \delta)\rangle_\delta$ is independent of $\Delta$. This confirms the scaling proposed in equation~\eqref{eqtime}. \\




We now aim to verify that an initial large separation of scales, namely for which $d_0/d_h\ll 1$, indeed leads to a universal $\mathcal{N}(d)\propto d^{-3/2}$ in the sub-Hinze scale regime. We will analyze laboratory and numerical data of bubble size distribution under breaking waves from previous work \citep{Deane_2002,Mostert_2021}. On top of this, we introduce a more idealized configuration consisting of a single large bubble injected in a turbulent flow, both numerically \cite{Riviere2021} and experimentally.

\begin{figure*}
\centering
\includegraphics[width=0.9\textwidth]{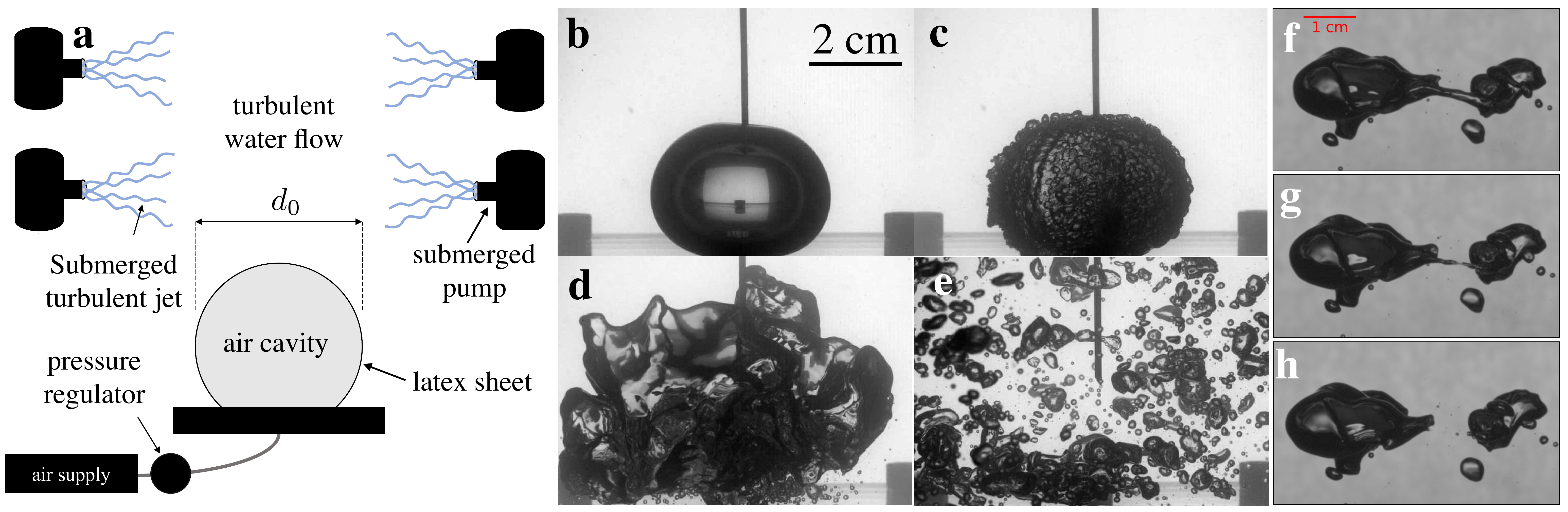}
\caption{({\bf a}) Sketch of the experimental set-up. (b-e) Successive snapshots of the release of a large bubble into a turbulent background flow. {\bf b}) Before the crack opening, an air pocket is trapped within a extended thins rubber sheet.  {\bf c}) Just after the membrane piercing, the membrane moves away, generating locally a high shear situation, but on a timescale much shorter than the typical turbulence time at the bubble scale (0.2 ms). The small wavelength disturbances are then dissipated by viscosity while the interface deforms at larger scale under the action of the background turbulence ({\bf d}). Eventually the bubble interface experiences multi breaking events, generating a broad distribution of bubble size ({\bf e}). ({\bf f-h}) Zoom in view of a typical breaking dynamics of a gas filament during the process. The time between images is 2ms.}
\label{presentation}
\end{figure*}




	We design an experiment that injects a unique bubble of initial size much larger than the Hinze scale $d_h$. Using a thin latex membrane, we pressurise an underwater air cavity of diameter $d_0 = 40$~mm, as shown in figure \ref{presentation}a. In the water phase, a turbulent flow is generated in an horizontal middle plane located above the initial air pocket, similarly to \cite{Ruth_PNAS_2019}. It is done by arranging and running four pumps pointing toward the center. The resulting velocity field is characterized using a Particle Image Velocimetry algorithm~\cite{Thielicke_2014}, which gives $u' = 0.25$~m/s, $L_\text{int} = 1.5$~cm, {$\epsilon = 0.7$}~m$^2$.s$^{-3}$ and $Re_\lambda = 340 \pm 40$, where $u'$ is the root mean squared (rms) velocity, $L_\text{int}$ the integral length scale and $Re_\lambda$ is the Taylor Reynolds number that characterizes the turbulent velocity fluctuations. These set the Hinze scale to $d_h \approx 4.8$~mm with $\gamma = 50$~mN/m and $\rho = 10^3$~kg/m$^3$. The ratio between the initial cavity diameter $d_0 = 40$~mm and the Hinze scale is therefore $d_0/d_h = 8.3$, which defines a Weber number $We = We_c (d_0/d_h)^{5/3} \approx 100$, corresponding to a large separation of scales. Note that in the literature $We_c$ varies between 1 and 5, depending on the details of the turbulence setup \cite{lalanne2019model,Lasheras1999,Risso1998}. We consider $We_c$ = 3 for consistency with the DNS \cite{Riviere2021}. The air pocket is released by piercing the membrane, which triggers a rapid crack opening. After a transient regime of interface deformations by interfacial instabilities, the bubble rises and deforms under the combined action of buoyancy and turbulent background flow. A comparison with the quiescent case shows that the bubble fragments are mainly produced by the turbulent background flow. {In the main turbulent region located between the four pumps, a broad range of bubble sizes is eventually generated, as illustrated by the successive snapshots of fig. \ref{presentation}b-e. The large air bubbles are highly deformed and lead to the formation of tiny air filaments, breaking down in small bubbles as illustrated in fig. \ref{presentation}f-h. To measure the size distribution quantitatively, we move the pumps 20 cm above the air pocket. We then process images taken with a high speed camera with a resolution of 15$\mu$m per pixel filming at 1000 fps to compute the distribution of bubble sizes in the region of the most intense turbulence. We compute the bubble size distribution averaged over 2 runs and $1.2$~s $\approx 20\,T_\text{int}$ each after the first break-up, with $T_\text{int} = L_\text{int}/u'$ the integral time scale associated to the correlation time of the largest eddies.
	
	Figure~\ref{distrib} shows the bubble size distributions $\mathcal{N}(d/d_h)$ obtained under breaking waves \citep{Deane_2002,Mostert_2021} and from a single large bubble breaking in a turbulent flow, both experimentally and numerically. Within all four data sets, for $d>d_h$, the distributions exhibit $\mathcal{N}(d) \propto d^{-10/3}$ scaling, in agreement with other previous experimental measures below a breaking wave~\citep{Loewen_1996,Rojas_2007,blenkinsopp2010} and in agreement with the classic break-up cascade argument from \citet{Garrett2000}. 
	  
	  For $d<d_h$, all four dataset also exhibit $\mathcal{N}(d) \propto d^{-3/2}$ accross a large range of scales. The bubble size distribution measured under breaking waves is in close agreement with the data obtained from single bubble break-up in turbulence, suggesting that the same underlying mechanisms are at play for the sub-Hinze bubble production outside if laboratory experiments. \al{This result also justifies that $F(\Delta, \delta)=F(\Delta)$. Indeed, since $\mathcal{N}(d) \propto d^{-3/2}$, the left hand side of equation \eqref{budget} is proportional to $d^{-3/2}$ and since, from the time analysis the $d$-dependency of the right hand side is $d^{-3/2}F(\Delta, \delta)$, one gets {\it a posteriori} that $F(\Delta, \delta)$ must be independent of $\delta$.}
	  
	The observed size distribution, considered alongside the mechanism we present for sub-Hinze bubble production, suggests that for these experimental cases, the production rate of sub-Hinze scale bubbles	is controlled by surface tension, through the breaking dynamics of gas filaments. It extends to sub-Hinze bubble production the framework of~\citet{Villermaux_2020}, who stated that for liquids, ligaments may universally control fragmentation processes. Contrary to many fragmentation processes in which a physical length scale sets the average fragmentation size, there is no such specific length scale, and a power law distribution is observed instead. These capillary effects only dominate the production of sub-Hinze bubbles, since for larger bubbles, the dynamics, and thereby the lifetime, of the parent bubbles can also be controlled by the eddy turnover time.
	

\begin{figure}
\centering
\includegraphics[width=0.6\columnwidth]{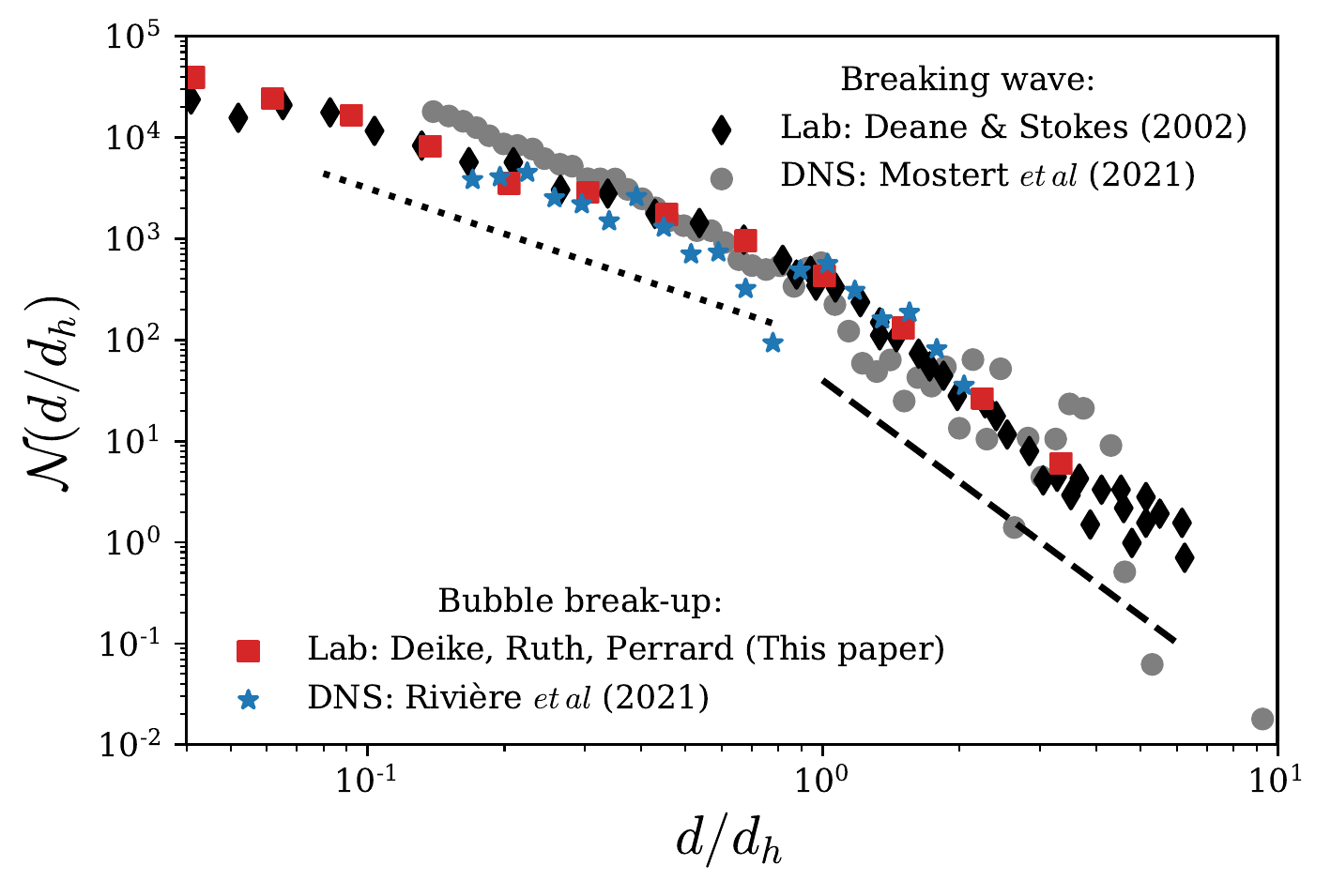}
\caption{Bubble size distribution obtained both experimentally and numerically in two different geometries: a breaking wave ($\blacklozenge$, \textcolor{gray}{$\bullet$}) and a single bubble breaking (\textcolor{red}{$\blacksquare$}, \textcolor{blue}{$\bigstar$}). The distribution exhibits two power laws: for $d>d_h$, $\mathcal{N}(d) \propto d^{-10/3}$ (dashed line), while $\mathcal{N}(d) \propto d^{-3/2}$ for $d<d_h$ (dotted line).}
\label{distrib}
\end{figure}	

The prefactor for the sub-Hinze distribution can eventually be evaluated using the continuity of $\mN$ at $d_h$, and the known expression for $d>d_h$ \cite{Deane_2002}, which gives:
\begin{align}
&\mN(d, t) = Q\epsilon^{-1/3}d^{-10/3} & \text{for }d>d_h \label{distribsuperHinze}\\
&\mN(d, t) = Q \left ( \frac{\We_c}{2} \frac{\gamma}{\rho} \right )^{-11/10}\epsilon^{2/5}d^{-3/2} & \text{for }d<d_h\label{distribsubHinze}
\end{align}
where $Q$ is the volume of air injected to the breaking cascade per volume of water per second.

In summary, when $d_0 \gg d_h$, large-scale inertial break-ups and small-scale capillary splitting events occur concurrently. The background turbulence sets the geometry of each break-up event over a time $t_c(\Delta)$ and then freezes relative to the capillary time scale~\cite{Ruth_PNAS_2019}, over which a cascade of small-scale splitting events occur. The classic turbulent inertial break-up scenario (eq.~\eqref{distribsuperHinze}) combined with the capillary driven fragmentation regime (eq.~\eqref{distribsubHinze}) provides a physical explanation for the entirety of the bubble size distribution when large air cavities break apart under the action of turbulence.

\begin{acknowledgments}
This work was supported by the NSF CAREER award 1844932 to L.D. A.R. was supported by an International Fund grant from Princeton University to L.D. S.P. and A.R. were supported by the Labex ENS-ICFP.  Computations were performed on the Princeton supercomputer Tiger2, as well as on Stampede, through XSEDE allocations to L.D. and W.M., XSEDE is an NSF funded program 1548562. We would like to acknowledge high-performance computing support from Cheyenne (doi:10.5065/D6RX99HX) provided by NCAR's Computational and Information Systems Laboratory, sponsored by the National Science Foundation.
\end{acknowledgments}





\bibliography{biblio.bib}

\end{document}